\documentstyle[twoside,fleqn,epsfig,espcrc2,amssymb]{article}

\makeatletter
\long\def\@makefigurecaption#1#2{\vskip 6mm #1. #2\par}
\long\def\@maketablecaption#1#2{\hbox to \hsize{\parbox[t]{\hsize}
  {#1 \\ #2}}\vskip 0.15ex}
\textfloatsep=\floatsep
\intextsep=\floatsep
\makeatother 

\newcommand{\al}{\alpha}
\newcommand{\be}{\beta}
\newcommand{\ga}{\gamma}

\newcommand{\de}{\delta}

\newcommand{\ep}{\varepsilon}

\newcommand{\La}{\Lambda}

\newcommand{\Si}{\Sigma}

\renewcommand{\>}{\rangle} 

\newcommand{\dsp}{\displaystyle}

\newcommand{\beq}{\begin{equation}}
\newcommand{\eeq}{\end{equation}}
\newcommand{\ba}{\begin{array}}
\newcommand{\bea}{\begin{eqnarray}}
\newcommand{\ea}{\end{array}}
\newcommand{\eea}{\end{eqnarray}}

\newcommand\comment[1]{ \hbox{[{\it Comment suppressed here.}\/]} }
\newcommand\hide[1]{}

\def\scr{\scriptstyle}

\newcommand{\skipover}[1]{}

\newcommand{\MeV}{{\rm MeV}}

\newcommand{\ms}{m_s}

\newcommand{\tsc}{2SC}
\newcommand{\cfl}{CFL}

\newcommand{\AmS}{{\protect\the\textfont2
  A\kern-.1667em\lower.5ex\hbox{M}\kern-.125emS}}

\title{High-density QCD: the effects of strangeness}
 
\author{Mark Alford\address{Center for Theoretical Physics,
Massachusetts Institute of Technology, Cambridge, MA 02139, USA}%
}

\begin{document}

\begin{abstract}
I discuss the zero temperature phase diagram of QCD, as a function of
baryon density and strange quark mass.  
The noteworthy points are that at sufficiently high density
chiral symmetry is always restored, and at low strange quark mass
there need be no phase transition between nuclear matter and quark matter.
I comment on the possibility that introducing a strange quark may make
it easier to see finite-density physics on the lattice.
\end{abstract}

\maketitle

\section{Introduction}
\label{sec:int}
Although lattice gauge theory has been successfully
applied to QCD at zero baryon density and non-zero temperature,
we know very little about QCD
at high density and low temperature, 
a regime which is physically relevant
to neutron star physics and low-energy heavy-ion collisions.
Strong evidence has been marshalled 
\cite{BarroisBailin,ARW2,RappETC}
that the ground state in this regime spontaneously breaks the color
gauge symmetry by a condensate of ``Cooper pairs'' of quarks.
The pattern of symmetry breaking has been found to be very different
for the cases of 2 and 3 flavors.
In this paper I report on the results of recent investigation into
the more realistic 2+1 flavor theory.

Let us start by reviewing the two- and three-flavor cases. \\
%
(1) Two masssless flavors.
The ground state is two-flavor color superconducting (\tsc):
chirally symmetric $u$-$d$ pairing \cite{ARW2,RappETC}.
The pattern is $\<q^\al_i C \ga_5 q^\be_j\> \sim \ep^{\al\be 3} \ep_{ij}$
(color indices $\al,\be$, flavor indices $i,j$), which breaks
$SU(3)_{\rm color} \to SU(2)_{\rm color}$, leaving
$SU(2)_L \times SU(2)_R$ unbroken. \\
(2) Three massless flavors. The ground state is 
3-flavor color-flavor locked (\cfl), with pairing between all flavors.
Chiral symmetry is broken\cite{ARW3}.
The pattern is 
$\<q^\al_i  C \ga_5 q^\be_j\>  \sim \de^\al_i \de^\be_j
- c\, \de^\al_j \de^\be_i$, breaking
$SU(3)_{\rm color}\times SU(3)_L \times SU(3)_R \to
SU(3)_{\rm color+L+R}$.
The ansatz is symmetric only under equal and opposite color and flavor
rotations. Since color is vectorial, this breaks the axial flavor symmetry.
Even though it only pairs left-handed quarks
with left-handed and right-handed with right-handed,
color-flavor locking {\em breaks} chiral symmetry. \\
(3) 2+1 flavors.
Even if one quark is massive, there is still 
a \cfl\ phase.
The pattern is more complicated than the
3-flavor case (see \cite{ABR} for details), but the essence is that
$u$-$s$ and $u$-$d$ pairing breaks the $SU(3)_{\rm color}\times
SU(2)_L\times SU(2)_R$ to $SU(2)_{\rm color+L+R}$, breaking chiral symmetry
because the flavor symmetries are locked to color.

\section{Phase diagram}

\begin{figure}[thb]
\epsfig{file=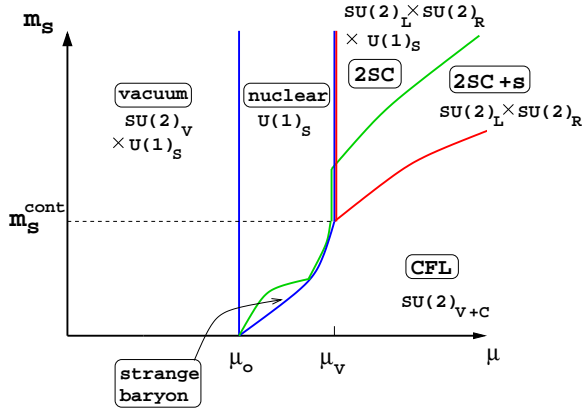,width=3in}
\caption{
Conjectured phase diagram for 2+1 flavor QCD at zero temperature
as a function of chemical potential $\mu$ and strange quark mass $\ms$.
The global symmetries of each phase are labelled.
} 
\label{fig:phasediagram}
\end{figure}

In Fig.~\ref{fig:phasediagram} we give a conjectured phase diagram for
2+1 flavor QCD, classifying the phases according to which global
symmetries of the Hamiltonian they leave unbroken.  These are the
$SU(2)_L\times SU(2)_R$ flavor rotations of the light quarks, and the
$U(1)_S$ of strangeness.

We make the following assumptions about the quark phases:
(1)~$m_u = m_d = 0$.
Including small $u,d$ masses would have little effect on pairing
\cite{BergesRajagopal,PisarskiRischke1OPT}.
(2)~Zero temperature.
(3)~The quark phase can be described by 
an NJL-type model, with a 4-fermion interaction
normalized by the zero-density chiral condensate.
(4)~Electromagnetism is ignored.
(5)~Weak interactions are ignored.

In the baryonic phases, we assume that baryon Fermi surfaces
are unstable to pairing in channels which preserve rotational invariance,
breaking internal symmetries such as isospin if necessary.

To explain Fig.~\ref{fig:phasediagram}, imagine following two lines
of increasing density ($\mu$), one at high $\ms$, then one at low $\ms$.

At high $\ms$, we start in the vacuum, where chiral symmetry is broken.
At $\mu_{\rm o}\sim 300~\MeV$, one finds nuclear matter, in which
$p$-$p$ and $n$-$n$ pairing breaks isospin. 
At $\mu_{\rm V}$, we find a first-order phase transition
to the \tsc\ phase  of color-superconducting
quark matter, in which the
red and green $u$ and $d$ quarks pair in isosinglet channels.
Pairing of the blue quarks is weak \cite{ARW2}, and we ignore it.
When $\mu$
exceeds the constituent strange quark mass we enter ``2SC+s''
in which there is a strange quark Fermi surface, with weak
$s$-$s$ pairing \cite{ABR}, but no $u$-$s$ or $d$-$s$ pairing.
Finally, when the chemical potential is high enough
that the Fermi momenta for the strange and light quarks
become comparable, there is a  first-order phase
transition to the color-flavor locked (\cfl) phase.
Chiral symmetry is once again broken.
Gapless superconductivity \cite{gapless} occurs in the metastable
region near the locking transition.

\begin{table}[htb]
\setlength{\arraycolsep}{0em}
\setlength{\tabcolsep}{0.5em}
\def\st{ \rule[-3.7ex]{0em}{8.4ex}}
\def\sst{\rule[-1.5ex]{0em}{4ex}}
\begin{tabular}{lcr|ccr} 
\hline
\sst Quark & \makebox[1.5em][r]{$\dsp \ba{c}SU(2)\\{\scr C+V}\ea$} & $Q'$ & 
   Hadron & \makebox[2em][c]{$SU(2)_{V}$}   & $Q$  \\
\hline
\st $\left(\ba{c} {bu}\\[1ex] {bd} \ea\right)$ & 
{\bf 2} &
$\ba{r} +1\\[1ex] 0 \ea$ &
$\left(\ba{c} p\\[1ex] n \ea\right)$  & 
{\bf 2} &
$\ba{r} +1\\[1ex] 0 \ea$ \\
\st $\left(\ba{c} {gs}\\[1ex] {rs} \ea\right)$ & 
{\bf 2} &
$\ba{r} 0\\[1ex]-1 \ea$ &
$\left(\ba{c} \Xi^0\! \\[1ex] \Xi^-\!\! \ea\right)$ & 
{\bf 2} &
$\ba{r} 0\\[1ex]-1 \ea$ \\
\hline
\sst $\left(\ba{c} {ru}- {gd}\\[1ex] 
 {gu}\\[1ex] {rd} \ea\right)$ & 
{\bf 3} &
$\ba{r} 0\\[1ex]+1 \\[1ex]-1\ea$ &
$\left(\ba{c} \Si^0 \\[1ex] \Si^+ \\[1ex] \Si^- \ea\right)$ & 
{\bf 3} &
$\ba{r} 0\\[1ex]+1 \\[1ex]-1\ea$ \\
\hline
$\dsp \ba{c}{ru}+{gd}\\ +\xi_- {bs}\ea$ &  {\bf 1} & 
$\ba{r}0\ea$ & 
 $\La$ & {\bf 1} &  $\ba{r}0\ea$ \\
\hline
\sst $\dsp \ba{c}{ru}+{gd}\\ -\xi_+ {bs}\ea$ & {\bf 1} & 
$\ba{r}0\ea$   &
  --- &  \\
\hline
\end{tabular}
\label{tab:qm}
\caption{The mapping between quark and hadronic states in the
\cfl\ phase.}
\end{table}

At low $\ms$, the story starts out the same way.
As density rises we enter
the nuclear matter phase with $pp$ and $nn$ pairing.
Then we enter the strange baryonic matter phase, with
Fermi surfaces for the $\La$ and/or $\Si$ and $\Xi$. 
These pair with themselves in
spin singlets, breaking $U(1)_S$ and isospin and chirality. 
We can imagine two possibilities for what happens next as $\mu$ increases
further.
(1) Deconfinement: the baryonic Fermi surface is replaced by
$u,d,s$ quark Fermi surfaces, which are unstable against pairing, and
we enter the CFL phase, described above. An ``isospin''
$SU(2)_{{\rm color}+V}$
is restored, but chiral symmetry remains broken.
(2) No deconfinement: the Fermi momenta of all of the octet
baryons are now similar enough that baryons with
differing strangeness can pair
in isosinglets
($p\Xi^-$, $n\Xi^0$, $\Si^+ \Si^-$, $\Si^0\Si^0$, $\La \La$),
restoring isospin.
The interesting point is that scenario (1) and scenario (2) are 
indistinguishable.
They both have a global $SU(2)$ symmetry, and an unbroken $U(1)$
gauge symmetry. This is the
``continuity of quark and hadron matter'' described by Sch\"afer and
Wilczek \cite{SchaeferWilczek}.  We conclude that for low enough
strange quark mass, $\ms<\ms^{\rm cont}$, there may be a region where
sufficiently dense baryonic matter has the same symmetries as quark
matter, and there need not be any phase transition between them.
The mapping between the baryonic and quark gaps is given in
table \ref{tab:qm}, along with their transformation properties
under the unbroken symmetries.

\section{Possible lattice calculations}

\begin{figure}[htb]
\begin{center}
\epsfig{file=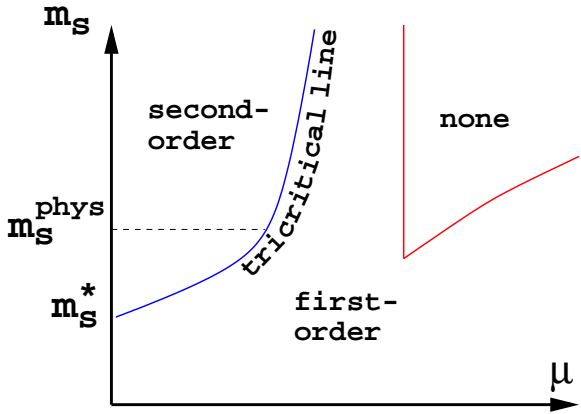,width=3in}
\end{center}
\caption{The order of the finite-temperature
chiral phase transition of QCD for $m_{ud}=0$.
The wedge at top right corresponds to the \tsc\ phase,
in which the ground state respects chiral symmetry, so there
is no chiral-restoring transition as $T$ rises.
If $m_{ud}$ is non-zero, the second-order region becomes a crossover,
and the tricritical line becomes a critical line.
}
\label{fig:tricritical}
\end{figure} 

It has been pointed out \cite{BergesRajagopal} that there is a
tricritical point in the $\mu$-$T$ plane for two-flavor QCD, which may
be experimentally detectable in heavy-ion collisions \cite{SRS}.  As
$\ms$ is reduced below its
physical value $\ms^{\rm phys}$, 
this tricritical point moves to lower chemical
potential, and at strange quark mass $\ms^*$ it occurs at
$\mu=0$ \cite{mscrit} (Fig.~\ref{fig:tricritical}).  
For lower $\ms$ the phase transition is first-order. Lattice calculations
give $\ms^*\approx \frac{1}{3}\ms^{\rm phys}$ \cite{jlqcd}.

For $\ms$ just above $\ms^*$, $\mu_{\rm crit}$ is very low, so one
should be able to estimate it by calculating derivatives of
observables such as the chiral susceptibility
with respect to $\mu$ at $\mu=0$. (Previously, $\mu$-derivatives
have only been calculated at $\ms=\infty$, where there is no 
nearby critical behavior, and no clear signal was
seen \cite{qcd-taro}). This could be extrapolated to
give some indication
of $\mu_{\rm crit}(\ms^{\rm phys})$, the value of
the critical chemical potential in real-world
QCD---a prediction that would be of direct value to heavy-ion
experimentalists, who need to know at what energy to run their
accelerators in order to see the phenomena predicted
to occur near the critical point \cite{SRS}.
 
It should also be noted that existing finite-$\mu$ techniques such as
imaginary chemical potential \cite{imag} may well be useful in this
crossover/critical region, since baryon masses become light as the
baryons continuously melt into quarks, so there will be little
suppression of the amplitudes of higher Fourier modes in imaginary
$\mu$.

\smallskip
\begin{center}
Acknowledgments
\end{center}%
I thank J. Berges, K. Rajagopal, and F. Wilczek for their
collaboration on the work reported here.
It was supported in part  by the U.S. Department
of Energy (D.O.E.) under cooperative research agreement \#DF-FC02-94ER40818.

\end{document}